# Pair creation supernovae at low and high redshift

N. Langer[1], C. A. Norman[2,3], A. de Koter[4], J. Vink[5], M. Cantiello[1] and S.-C. Yoon[6]

[1] Astronomical Institute, Utrecht University, Princetonplein 5, 3584 CC, Utrecht, The Netherlands
[2] The Johns Hopkins University, Homewood Campus, Baltimore, MD 21218
[3] Space Telescope Science Institute, 3700 San Martin Drive, Baltimore, MD 21218
[4] Astronomical Institute *Anton Pannekoek*, University of Amsterdam, Netherlands
[5] Armagh Observatory, College Hill, Armagh BT61 9DG, NI, UK
[6] Department of Astronomy and Astrophysics, University of California, Santa Cruz, CA 95064, USA

**ABSTRACT**

*Context.* -
*Aims.* Pair creation supernovae (PCSN) are thought to be produced from very massive low metallicity stars. The spectacularly bright SN 2006gy does show signatures expected from PCSNe. Here, we investigate the metallicity threshold below which PCSN can form and estimate their occurrence rate.
*Methods.* We perform stellar evolution calculations for stars of 150 M$_\odot$ and 250 M$_\odot$ of low metallicity (Z$_\odot$/5 and Z$_\odot$/20), and analyze their mass loss rates.
*Results.* We find that the bifurcation between quasi-chemically homogeneous evolution for fast rotation and conventional evolution for slower rotation, which has been found earlier for massive low metallicity stars, persists in the mass range considered here. Consequently, there are two separate PCSN progenitor types: (I) Fast rotators produce PCSNe from very massive Wolf-Rayet stars, and (II) Slower rotators that generate PCSNe in hydrogen-rich massive yellow hypergiants.
*Conclusions.* We find that hydrogen-rich PCSNe could occur at metallicities as high as Z$_\odot$/3, which — assuming standard IMFs are still valid to estimate their birth rates — results in a rate of about one PCSN per 1000 supernovae in the local universe, and one PCSN per 100 supernovae at a redshift of $z = 5$. PCSNe from WC-type Wolf-Rayet stars are restricted to much lower metallicity.

## 1. Introduction

Pair creation supernovae (PCSNe) are thought to be produced by stars which are, and remain throughout their lives, very massive. As they are radiation pressure dominated, the electron-positron pair production occurring in their cores for temperatures in excess of $\sim 10^9\,K$ can decrease the adiabatic index below 4/3 and destabilize the core (Fowler & Hoyle 1964, Kippenhahn & Weigert 1990). While the most massive stars ($\gtrsim 260\,M_\odot$) are thought to collapse into black holes (Bond et al. 1984, Heger & Woosley 2002), the ensuing explosive oxygen burning may disrupt stars with initial masses in the range of $\sim 100\,M_\odot ... 260\,M_\odot$ and thus produce a pair creation supernova (Ober et al. 1983, El Eid & Langer 1986, Heger & Woosley 2002).

PCSNe have mostly been considered in the context of pregalactic (Pop III) stars (Ober et al. 1983, Heger & Woosley 2002), since locally very massive stars are thought to lose mass at a high rate: a Galactic $120\,M_\odot$ star is expected to end as a Wolf-Rayet star with a mass of the order of $10\,M_\odot$ (Meynet & Maeder 2005). Heger et al. (2003) pointed out that there is a finite metallicity threshold below which PCSNe would occur, due to the strong metallicity dependence of massive star winds. However, this metallicity threshold has not been investigated in detail. The interest in this has been triggered by the recent supernova 2006gy, the brightest supernova which was ever found. Its properties might well correspond to a PCSN: its extreme brightness could relate to a large radius in combination with a high nickel mass (Smith et al. 2007, Scannapieco et al. 2005), and its slow evolution and expansion velocity correspond well to PCSN explosion models (Heger & Woosley 2002, Scannapieco et al. 2005). This raises the question of the likelihood of PCSNe occurring in the local universe. This question is intrinsically interesting, independent of the particular event of SN 2006gy.

To this end, we investigate the metallicity threshold for the occurrence of PCSNe. We perform stellar evolution calculations into the PCSN regime, to investigate the possible range of properties of PCSN progenitor stars. As the fate of potential PCSN progenitors depends mostly on stellar mass loss rates, we then discuss the relevant mass loss rates and derive metallicity thresholds for the main branches of PCSNe. Finally, we estimate the occurrence rate of PCSNe, and their progenitor stars, in the local universe and at high redshift.

## 2. Stellar evolution models

We have computed several stellar evolution sequences for initial masses of $150\,M_\odot$ and $250\,M_\odot$ at low-metallicity (Table 1). All physical ingredients and assumptions in these calculations are identical to the ones by Yoon et al. (2006). In particular, the mass loss recipe of Kudritzki et al. (1989) with a metallicity scaling proportional to $(Z/Z_\odot)^{0.69}$ (Vink et al. 2001) was used to compute the mass loss rate of main sequence stars. For Wolf-Rayet phases, mass loss was computed according to Eq. (1) of Yoon et al., which includes mass loss enhancement for metal-

*Send offprint requests to*: N. Langer

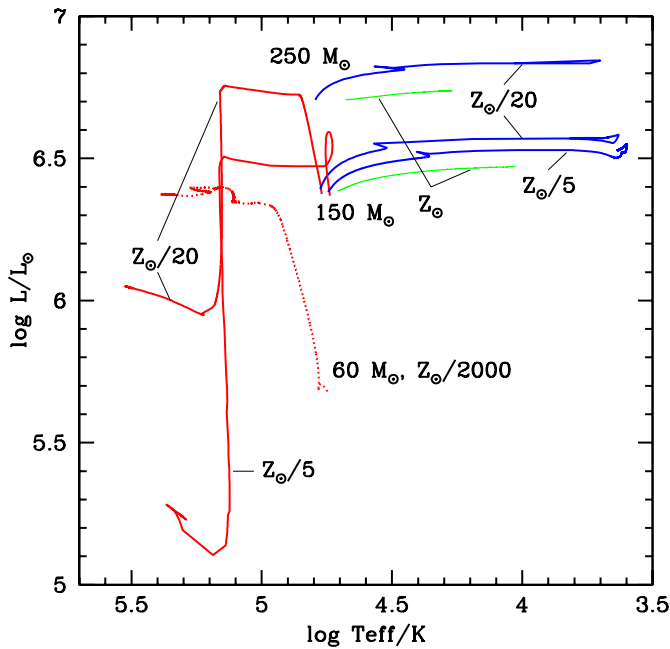

**Fig. 1.** Evolutionary tracks of the computed sequences (cf. Table 1) in the HR diagram, from the zero age main sequence to their pre-SN position. The two sequences which rotate rapidly initially (150 M$_\odot$; red lines in the electronic version) evolve quasi-chemically homogeneously, i.e. blueward, and end their evolution with an iron core collapse. The slow rotators (150 M$_\odot$ and 250 M$_\odot$; blue lines) become yellow hypergiants after the main sequence evolution. Additionally, the 60 M$_\odot$ rapid rotator ($v_{\rm rot,i}/v_{\rm Kepler} = 0.3$) for $Z = 10^{-5}$ from Yoon et al. (2006) is shown (red). Furthermore, solar metallicity tracks for 150 M$_\odot$ and 250 M$_\odot$ from Figer et al. (1998) are shown (green lines); they end during core hydrogen burning, when their surfaces become unstable.

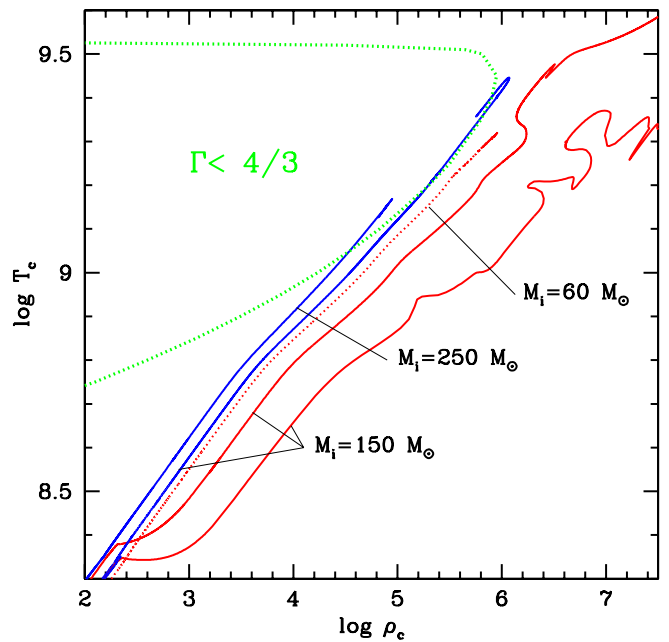

**Fig. 2.** Evolutionary tracks of our stars in the $\log \rho_c - \log T_c$-diagram. The region of pair-instability is indicated ($\Gamma < 4/3$). The red lines correspond to our initially rapidly rotating 150 M$_\odot$ models for $Z=Z_\odot/5$ (lower curve) and $Z=Z_{M_\odot}/20$ (upper curve). Shown in blue are the slowly rotating 150 M$_\odot$ model at $Z=Z_\odot/20$, and the 250 M$_\odot$ model at $Z=Z_\odot/20$.

enriched surfaces, and a metallicity scaling as proposed by Vink & de Koter (2005). For cool stars, we used the mass loss rate of Nieuwenhuijzen & de Jager (1990) with a metallicity scaling of $(Z/Z_\odot)^{0.50}$.

All sequences have been computed including the physics of rotation and rotationally induced mixing and magnetic fields, as in Yoon et al. However, we computed models for slow rotation, i.e. with an initial equatorial rotation rate of 10 km s$^{-1}$, and for fast rotation, adopting rotation at 500 km s$^{-1}$ or about 40% of Kelperian rotation. Key properties of all computed sequences are listed in Table 1.

As for the low-metallicity stars of lower mass computed by Yoon et al. (2006), we find the rapid rotators to evolve quasi-chemically homogeneous. This is to be expected, since rotationally induced mixing is faster in more massive stars, where gas pressure becomes less and radiation pressure more important. Yoon at al. found quasi-chemically homogeneous evolution for $v_{\rm rot,i}/v_{\rm Kepler} > 0.2$ at 60 M$_\odot$, which was their largest considered mass. Our fast rotating 150 M$_\odot$ models (Sequences No. 2 and 4; cf. Tab. 1) evolve quickly into Wolf-Rayet stars (Fig. 1), and lose large amounts of mass and angular momentum; they end at 7 M$_\odot$ ($Z/Z_\odot=0.2$) and 23 M$_\odot$ ($Z/Z_\odot=0.05$). Due to their reduced mass, they avoid the pair creation instability regime and undergo stable oxygen burning (Fig. 2), evolving towards iron core collapse. We also show the trajectory of a rapidly rotation ($v_{\rm rot,i}/v_{\rm Kepler} = 0.3$) 60 M$_\odot$ model at $Z = 10^{-5}$ of Yoon et al. (2006) in Figs. 1 and 2, to illustrate that, if strong mass loss is avoided — i.e. for low enough metallicity —, the chemically ho-mogeneous models may also develop into PCSNe (cf. Sect. 3.2). This model ends with a total mass of about 51 M$_\odot$.

As shown in Fig. 1, our slowly rotating models (Seq. No. 1, 3, 5) pursue more conventional tracks in the HR diagram and evolve redward after core hydrogen burning. The total amount of mass lost during core hydrogen burning is rather moderate for those models (Tab. 1); their surfaces do not become enriched by hydrogen burning products. After core hydrogen exhaustion, these stars evolve into cool supergiants. None of the models becomes cool enough to form dust; their effective temperatures remain higher than 4 200 K (cf. also Maeder & Meynet 2001). The lower metallicity models of 150 M$_\odot$ and 250 M$_\odot$ (Seq. No. 3 and 5) suffer significant mass loss during core helium burning. However, the helium core of both stars is not affected by the mass loss, and both models evolve into the pair creation instability and start collapsing before core oxygen ignition (Fig. 2)

According to our mass loss prescription, the 150 M$_\odot$ sequence at $Z/Z_\odot=0.2$ loses about all of its hydrogen-rich envelope during core helium burning. Numerical problems prevented us from computing this model to core helium exhaustion. An extrapolation of the mass loss rate during the core helium burning stage up to core helium exhaustion leads to expect a total mass loss during core helium burning of the order of 90 M$_\odot$; this would bring the star onto the Wolf-Rayet branch and into the core collapse regime. However, we can not exclude that star would settle in the blue supergiant regime where mass loss is smaller. Therefore, the fate of this model remains uncertain.

## 3. PCSN metallicity thresholds

Most important for estimating the metallicity thresholds for PCSNe are the mass loss rates of very massive stars during their evolution. From the evolutionary models shown above, it is clear that for the rapid rotators, the Wolf-Rayet mass loss is detrimen-tal. For slow rotators it is the mass loss on the main sequence, and in the cool supergiant stage. A problem in evaluating this quantitatively is that the required mass loss rates are not accessible observationally, and that theoretical estimates have to be stretched quite far to yield numbers in our luminosity and metallicity range. In the following, we discuss mass loss for slowly and fast rotating models, separately.

**Table 1.** Key properties of the computed sequences: initial mass, metallicity and rotation rate, total main sequence and post main sequence mass loss, final total, He-core and CO-core mass, pre-SN luminosity, effective temperature, and radius, central temperature of last computed model, and fate (cc=iron core collapse, PCSN=pair creation supernova).

| # | $M_i$ | $Z$ | $v_{\rm rot,i}$ | $\Delta M_{\rm MS}$ | $\Delta M_{\rm PMS}$ | $M_{\rm f}$ | $M_{\rm He,f}$ | $M_{\rm CO,f}$ | $L_{\rm f}$ | $T_{\rm eff,f}$ | $R_{\rm f}$ | $T_{\rm c,end}$ | fate |
|---|---|---|---|---|---|---|---|---|---|---|---|---|---|
|   | $M_\odot$ | $Z_\odot$ | km s$^{-1}$ | $M_\odot$ | $M_\odot$ | $M_\odot$ | $M_\odot$ | $M_\odot$ | log $L/L_\odot$ | K | $R_\odot$ | $10^8$ K | |
| 1 | 150 | 0.2 | 10 | 14 | ~ 90 | ~ 45 | - | - | - | - | - | 2.27 | ? |
| 2 | 150 | 0.2 | 500 | 74 | 69 | 7 | 6.8 | 5.4 | 5.2 | 202 000 | 0.3 | 22.7 | cc |
| 3 | 150 | 0.05 | 10 | 4 | 53 | 93 | 71 | 64 | 6.5 | 4 200 | 3300 | 22.7 | PCSN |
| 4 | 150 | 0.05 | 500 | 31 | 96 | 23 | 23 | 22 | 6.0 | 315 000 | 0.4 | 50.1 | cc |
| 5 | 250 | 0.05 | 10 | 10 | 71 | 169 | 121 | 109 | 6.8 | 10 700 | 760 | 14.1 | PCSN |
|   | 60 | 0.0005 | 400 | 0.02 | 8.5 | 51 | 51 | 46 | 6.4 | 219 000 | 1.1 | 20.6 | cc |

**Table 2.** Estimated main sequence mass loss according to Vink et al. (2001), using $T_{\rm eff} = 40\,000$ K and $v_\infty/v_{\rm esc} = 2.6$. For the 150 $M_\odot$ case, we adopted log $L/L_\odot = 6.5$ and $\tau = 2.8$ Myr, while for the 250 $M_\odot$ case log $L/L_\odot = 6.8$ and $\tau = 2.5$ Myr.

|   | 150 $M_\odot$ | | 250 $M_\odot$ | |
|---|---|---|---|---|
| $Z$ | log $\dot M$ | $\Delta M$ | log $\dot M$ | $\Delta M$ |
| $Z_\odot$ | -4.46 | 96 | -4.10 | 200 |
| $Z_\odot/3$ | -4.87 | 38 | -4.51 | 77 |
| $Z_\odot/10$ | -5.32 | 13 | -4.96 | 27 |
| $Z_\odot/30$ | -5.73 | 5 | -5.37 | 11 |

### 3.1. Slow rotation

For our slowly rotating models, we see that the main sequence mass loss at $Z/Z_\odot=0.2$ is not critical for the evolution towards PCSNe; the corresponding 150 $M_\odot$ sequence predicts only 14 $M_\odot$ of mass loss. As a criterion to prevent PCSNe, one may require that a good fraction, say a third, of the total mass of the star would need to be lost. Using the Vink et al. (2001) mass loss recipe, one obtains a critical metallicity of about $Z_{\rm PCSN,MS} \simeq 1/3$ (cf. Tab. 2). This implies that e.g. at SMC metallicity, the main sequence winds would not prevent a very massive slow rotator from evolving towards a PCSN.

The post main sequence mass loss of slowly rotating potential PCSN progenitors, i.e. for yellow supergiants, is more difficult to assess. Consequently, none of the models we computed evolves into the red supergiant regime, where dust formation may boost the mass loss rate. However, one important question is whether the stars, on their way to cool temperatures, would become Luminous Blue Variables (LBVs) and suffer from eruptive mass loss. While a good theoretical understanding of the LBV phenomenon is still lacking, metal-rich stellar models of very massive stars do become unstable for effective temperatures significantly below 20 000 K, as these models have high enough Rosseland-mean opacities in their outermost layers to hit the Eddington limit (Figer et al. 1998; cf. Fig. 1). Our low metallicity models here do not encounter this problem (see also Maeder & Meynet 2001). We take this as an indication that an LBV-type mass loss can be avoided at low metallicity.

The yellow supergiant mass loss rate is very uncertain. The metallicity-scaled rate of Nieuwehuijzen & de Jager (1990) as used in our models implies a post-main sequence wind induced metallicity threshold between $Z/Z_\odot=0.2$ and $Z/Z_\odot=0.05$. However, the physics of the winds from yellow supergiants is not well understood. Schröder & Cuntz (2005) work out a mass loss rate based on the physics of thermally driven winds, in which case no metallicity dependence is expected. Schröder & Cuntz (2007) show that this mass loss rate fits observations well over the whole observationally accessible dwarf, giant and supergiant regime. If we apply it to our 150 $M_\odot$ models at $T_{\rm eff} = 4000$ K, we obtain a mass loss rate of $1.9\,10^{-5}$ $M_\odot$ yr$^{-1}$, or a total mass loss of 5.2 $M_\odot$ over the post main sequence life time. While there is a large uncertainty, this shows that currently the consideration of post main sequence mass loss can not be used to exclude PCSNe from slowly rotating massive stars with metallicities below about $Z/Z_\odot=1/3$. Interestingly, the pre-SN mass loss rate of SN 2006gy is constraint by the soft X-ray measurements to $1...5\,10^{-4}$ $M_\odot$ yr$^{-1}$, which, extrapolated over $3\,10^5$ yr would result in a mass loss of $30...140\,M_\odot$.

### 3.2. Fast rotation

The fast rotating PCSN progenitor candidates undergo quasi-chemically homogeneous evolution (Yoon & Langer 2005, Yoon et al. 2006, Woosley & Heger 2006), and therefore evolve into Wolf-Rayet stars already during core hydrogen burning. As the mass loss rates during the core hydrogen burning Wolf-Rayet stage is significantly higher than the mass loss rate of the slowly rotating counterparts (cf. Table 1), our rapidly rotating 150 $M_\odot$ star at $Z/Z_\odot=0.2$ loses already 74 $M_\odot$ during core hydrogen burning, which disqualifies it as a PCSN progenitor (cf. Fig. 2). However, during the core hydrogen burning Wolf-Rayet stage, photon scattering with iron ions provides the main force to drive the stellar wind, which is strongly decreasing at least down to metallicities of $Z/Z_\odot = 10^{-4}$ (Vink & de Koter 2005). Consequently, looking at our rapidly rotating models (Tab. 1), rapid rotators above a metallicity of about $Z/Z_\odot=0.05$ are excluded from evolving into PCSNe.

During the post-main sequence evolution, the surfaces of the rapid rotators are quickly enriched with carbon and oxygen, transforming them into WC-type Wolf-Rayet stars. This enhances the mass loss rate, especially for $Z/Z_\odot < 0.1$ (Vink & de Koter 2005). The final masses of our rapidly rotating 150 $M_\odot$ sequences are 7 $M_\odot$ and 23 $M_\odot$, indicating that the metallicity limit to allow for PCSNe from initially rapidly rotating stars is much smaller than $Z/Z_\odot=0.05$. Vink & de Koter (2005) predict that the WC-type mass loss rate reaches a lower limit at about $Z/Z_\odot = 10^{-4}$, and extrapolating their results to 150 $M_\odot$ and 250 $M_\odot$ predicts only about 4 $M_\odot$ and 9 $M_\odot$ being lost during the WC stage at $Z/Z_\odot = 10^{-4}$. Therefore, PCSNe are likely to occur from the lowest metallicity rapid rotators. This agrees with the fact that the rapidly rotating $60\,M_\odot$ models at $Z/Z_\odot = 5\,10^{-4}$ of Yoon et al. (2006) lose only about $10\,M_\odot$ and are actually very close to the pair-instability (cf. Figs. 1 and 2). We conclude that the metallicity threshold for PCSN from the rapidly rotating branch is roughly at $Z/Z_\odot = 10^{-3}$.

**Table 3.** PCSN-to-SN ratios at various redshifts $z$, for two different PCSN-metallicity thresholds $Z_{PCSN}$ (first 2 lines), and for PCSNe from rapid rotators with a metallicity threshold of $Z_\odot/1000$. Here $f_r$ is the fraction of rapid rotators; according to Yoon et al. (2006; Fig. 5), it might be $f_r \simeq 0.25$.

| $Z_{PCSN}$ | $z = 0$ | $z = 2$ | $z = 5$ |
|---|---|---|---|
| $Z_\odot/3$ | 0.001 | 0.004 | 0.01 |
| $Z_\odot/10$ | 0.0001 | 0.001 | 0.004 |
| $Z_\odot/1000$ | $4\,10^{-7} f_r$ | $10^{-6} f_r$ | $10^{-5} f_r$ |

## 4. Observational consequences

In the following, we want to discuss occurrence and detection rates of PCSNe at various redshifts. For this purpose, we assume that slowly rotating massive stars produce PCSNe for initial masses in the range $140\,M_\odot$ to $260\,M_\odot$ (Heger & Woosley 2002) and for $Z/Z_\odot < 1/3$, while rapid rotators do so for initial masses above $80\,M_\odot$ and for $Z/Z_\odot < 1/1000$. The uncertainties in the post-main sequence mass loss rates discussed above imply that this way we derive upper limits to the occurrence and detection rates.

The occurrence rate of PCSNe depends decisively on the formation rate of stars with masses above the mentioned limits. Not much is known about that, due to the strong decline of the initial mass function (IMF) with mass. Figer (2005) finds that in the Arches cluster close the the Galactic center, there is a firm upper limit to the IMF at $150\,M_\odot$. On the other hand, Kudritzki et al. (1996) give evidence for stars with masses of up to $200\,M_\odot$ in the LMC. On the other hand, PCSN initial masses could be lower than the numbers suggested by Heger & Woosley (2002), either due to rotational mixing (cf. Sect. 3.2) or due to efficient convective core overshooting (cf., Langer & El Eid 1986) which is not well constrained at the considered masses. In lieu of more meaningful constraints for their birthrate especially at low metallicity, we simply use the Salpeter IMF to estimate the birth rate of PCSN progenitors. This results in stars above $140\,M_\odot$ making up for about 1% of all stars above the supernova threshold of $8\,M_\odot$, while stars above $80\,M_\odot$ amount to about 4%.

### 4.1. Local universe

In the local universe, the ratio of the birth rate of massive stars with $Z/Z_\odot < 1/3$ to that of all massive stars is about $1/10$ (Langer & Norman 2006), which then results in a local PCSN/SN-ratio of about $1/1000$ (cf. Tab. 3). Most PCSNe locally are expected to occur in yellow hypergiants with massive hydrogen-rich envelopes, since PCSNe produced from initially rapidly rotating stars are expected to be negligible in the local universe. The large radii of these PCSN progenitors will lead to very bright events (Young 2004, Scannapieco et al. 2005). The slow expansion velocity of PCSN (Heger & Woosley 2002, Scannapieco et al. 2005) implies a supernova light curve with a broad maximum. All this appears to be consistent with SN2006gy (Ofek et al. 2007, Smith et al. 2007).

If SN2006gy were a typical PCSN, it being about 10 times brighter, and shining about twice as long as an average supernova (cf. Fig. 2 in Smith et al. 2007) would imply an increased detection probability of $2 \times 10^{3/2} \simeq 60$ for PCSNe. I.e., the ratio of the detection probability of a PCSN to that of an average supernova in a local magnitude-limited search would work out to be 0.06, a number which appears too high to account for only one observed case. Thus, either SN2006gy is not a typical PCSN, or the PCSN metallicity threshold is lower than $Z_\odot/3$.

In fact, Herzig et al. (1990) derived realistic light curves for PCSNe, and considering compact progenitor models for PCSNe at the lower mass limit they found a maximum bolometric brightness of only $M_{bol} \simeq -14.6$, compared to $M_{bol} \simeq -22$ for SN2006gy (Smith et al. 2007). This is consistent with only small amounts of $^{56}$Ni being produced in the lower third or so of the PCSN progenitor mass range according to Heger & Woosley (2002), while the nickel mass implied for SN2006gy is of the order of $20\,M_\odot$ (Smith et al. 2007), which would require a progenitor mass in the upper part of the PCSN progenitor mass range. On the other hand, our slowly rotating $150\,M_\odot$ model at $Z_\odot/20$ (Seq. #3) explodes with a very large radius ($3300\,R_\odot$; cf. Table 1), which may lead to a high peak brightness even without a large amount of radioactive energy input (Young 2004).

Scannapieco et al. (2005) compute light curves of population III PCSNe, and find a wide range of peak luminosities and light curve peak widths. Furthermore, the fraction of stars which rotates rapidly enough to undergo chemically homogeneous evolution (Yoon et al. 2006) does not contribute significantly to the number of PCSN in the local universe. Furthermore, supernova searches are not optimally designed to find bright, rare objects with long ($\sim 1 yr$) time variations. Thus, for a PCSN metallicity threshold of $Z_\odot/3$, the detection PCSN probability may be of the order 0.001/SN if their progenitors form according to a Salpeter IMF. Adopting a star formation rate in the local universe of $0.0063\,M_\odot\,yr^{-1}\,Mpc^{-1}$ (Bouwens et al. 2004) results in about one PCSN per year within a radius of 100 Mpc for $Z_{PCSN} = Z_\odot/3$, and one per 10 years for $Z_{PCSN} = Z_\odot/10$. We note that the metallicity of NGC 1260, the S0/Sa peculiar host of SN 2006gy, is measured from the Mg$_2$ index to be $\gtrsim Z_\odot/2$ (Ofek et al. 2007).

We want to point out that also the progenitors of PCSNe might be detectable in the local universe. They have luminosities in the range $\log L/L_\odot \simeq 6.4...6.9$ and since they have only small bolometric corrections this translates into $M_{bol} \simeq M_V \simeq -11.2... -12.5$. This makes them $13^{th}$ magnitude stars at 1 Mpc distance. Even the progenitor of SN 2006gy at a distance of 73 Mpc (Smith et al. 2006) might have been visible with $m_V \simeq 22$ mag, or with $m_V \simeq 24$ mag including two magnitudes extinction (Smith et al. 2007). For the local star formation rate quoted above, we expect about 200 000 yellow hypergiant progenitors for $Z_{PCSN} = Z_\odot/3$, and ten times less for $Z_{PCSN} = Z_\odot/10$, within a radius of 100 Mpc.

The Small Magellanic Cloud may be a local test case for the PCSN metallicity threshold. If its metallicity ($\sim Z_\odot/5$) is below the threshold, then from the number ratio of O to WR stars of 0.015 (with O stars being stars more massive than about $15\,M_\odot$), and about 10 WR stars (Azzopardi et al. 1988), we expect about 14 O stars stars above $140\,M_\odot$, and about one PCSN progenitor in the YSG phase, and a PCSN rate of the order of $10^{-6}\,yr^{-1}$. We believe that none of these numbers can exclude that the SMC metallicity is indeed below the PCSN threshold.

### 4.2. High z, low Z universe.

The formalism of Langer & Norman (2006) allows to compute the occurrence rates of PCSNe in the high redshift universe, for different metallicity thresholds. Tab. 3 shows that at redshift $z =$ 2, the PCSN/SN ratio is 4 or 10 times higher than locally, for $Z_{PCSN} = Z_\odot/3$ and $Z_{PCSN} = Z_\odot/10$, respectively. In the first case, the metallicity bias for PCSNe has vanished at redshift $z = 5$. The fraction of PCSNe from rapid rotators remains negligible even out to a redshift of $z = 5$. However, PCSNe from rapid rotators may play an important role in the Pop III era.

Finally, in environments with $Z < Z_{PCSN}$, PCSNe may significantly contribute to the nucleosynthesis. Even though PCSNe constitute only 1% of the supernovae in that case, they encompass 10% of the mass which star formation incorporates into supernova progenitors. And while a core collapse supernova produces from zero up to a few solar masses of metals, a PCSN liberates 50...120 $M_\odot$ of metals; i.e. PCSNe produce about half of all metals (Heger & Woosley 2002, Umeda & Nomoto 2002). Thus, the consideration of their yields may significantly constrain the number of PCSNe. Extensive PCSN yields have been computed so far only for $Z = 0$, and the lack of an odd-even pattern in intermediate mass elements of extremely metal-poor halo stars appears to question the existence of Population III PCSNe (Heger & Woosley 2002, Umeda & Nomoto 2005). Detailed nucleosynthesis calculations for for PCSNe at $Z \simeq Z_\odot/3$ are presently not available, but Heger & Woosley's zero metallicity PCSN models show a "surprising overall approximate agreement" with solar system abundances in the range oxygen to nickel. Ballero et al. (2006) find that the nucleosynthesis signature of Population III PCSNe can not affect the predicted abundance ratios for the evolution of the Milky Way, even in its earliest evolutionary phase. Therefore, it may be difficult to rule out a PCSN metallicity threshold as high as $Z \simeq Z_\odot/3$ on nucleosynthesis reasons at present. Metallicity dependent PCSN yields and subsequent chemical evolution modeling is needed to obtain more stringent constraints on $Z_{PCSN}$ from nucleosynthesis.

In general, detecting low redshift PCSNe and understanding their physical properties will be a significant help in designing observational studies to observe PCSNe with JWST at high redshift and very low metallicity.

*Acknowledgements.* This work benefited from discussions during the Lorentz-Center workshop "From massive stars to supernova remnants" in Leiden held in Aug. 2007. Support for this work was provided by the National Aeronautics and Space Administration through Chandra Award Number GO6-7130X issued by the Chandra X-ray Observatory Center, and by the National Aeronautics and Space Administration under Grant NNG04GQ04G issued through the Search for Origins Program.